\documentclass[12pt,twocolumn,a4paper]{aastex63}
\usepackage{amsmath}
\usepackage{natbib}
\usepackage[figuresright]{rotating}
\usepackage{float}

\def\beq{\begin{eqnarray}}
\def\eeq{\end{eqnarray}}
\def\azero{A0620$-$00}
\def\xray{\hbox{X-ray}}
\def\vsini{$v\sin i$}
\def\simgt{\lower 2pt \hbox{$\, \buildrel {\scriptstyle >}\over {\scriptstyle \sim}\,$}}
\def\simlt{\lower 2pt \hbox{$\, \buildrel {\scriptstyle <}\over {\scriptstyle \sim}\,$}}

\shorttitle{Disk Veiling of A0620$-$00}
	
\begin{document}

\title{The Disk Veiling Effect of the Black Hole Low-Mass X-ray Binary A0620$-$00\footnote{This paper includes data gathered with the 6.5 meter Magellan Telescopes located at Las Campanas Observatory, Chile.}}

\correspondingauthor{Jianfeng Wu}
\email{wujianfeng@xmu.edu.cn}

\author{Wan-Min Zheng}
\affiliation{Department of Astronomy, Xiamen University, Xiamen, Fujian 361005, China}

\author[0000-0003-4202-1232]{Qiaoya Wu}
\affiliation{Department of Astronomy, Xiamen University, Xiamen, Fujian 361005, China}
\affiliation{Department of Astronomy, University of Illinois at Urbana-Champaign, Urbana, IL 61801, USA}

\author[0000-0001-7349-4695]{Jianfeng Wu}
\affiliation{Department of Astronomy, Xiamen University, Xiamen, Fujian 361005, China}

\author[0000-0003-3116-5038]{Song Wang}
\affiliation{Key Laboratory of Optical Astronomy, National Astronomical Observatories, Chinese Academy of Sciences, Beijing 100101, China}

\author[0000-0002-0771-2153]{Mouyuan Sun}
\affiliation{Department of Astronomy, Xiamen University, Xiamen, Fujian 361005, China}

\author[0000-0001-9346-3677]{Jing Guo}
\affiliation{Department of Astronomy, Xiamen University, Xiamen, Fujian 361005, China}

\author[0000-0002-7600-1670]{Junhui Liu}
\affiliation{Department of Astronomy, Xiamen University, Xiamen, Fujian 361005, China}

\author[0000-0002-5839-6744]{Tuan Yi}
\affiliation{Department of Astronomy, Xiamen University, Xiamen, Fujian 361005, China}

\author[0000-0002-2419-6875]{Zhi-Xiang Zhang}
\affiliation{Department of Astronomy, Xiamen University, Xiamen, Fujian 361005, China}

\author[0000-0003-3137-1851]{Wei-Min Gu}
\affiliation{Department of Astronomy, Xiamen University, Xiamen, Fujian 361005, China}

\author[0000-0003-4874-0369]{Junfeng Wang}
\affiliation{Department of Astronomy, Xiamen University, Xiamen, Fujian 361005, China}

\author[0000-0003-3057-5860]{Lijun Gou}
\affiliation{Key Laboratory for Computational Astrophysics, National Astronomical Observatories, Chinese Academy of Sciences, Beijing 100101, China}
\affiliation{School of Astronomy and Space Sciences, University of Chinese Academy of Sciences, Beijing 100049, China}

\author{Jifeng Liu}
\affiliation{Key Laboratory of Optical Astronomy, National Astronomical Observatories, Chinese Academy of Sciences, Beijing 100101, China}
\affiliation{School of Astronomy and Space Sciences, University of Chinese Academy of Sciences, Beijing 100049, China}
\affiliation{WHU-NAOC Joint Center for Astronomy, Wuhan University, Wuhan, Hubei 430072, China}


\author{Paul J. Callanan}
\affiliation{Department of Physics, University College Cork, Cork, Ireland}

\author[0000-0001-6947-5846]{Luis C. Ho}
\affiliation{Kavli Institute for Astronomy and Astrophysics, Peking University, Beijing 100871, China}
\affiliation{Department of Astronomy, School of Physics, Peking University, Beijing 100871, China}

\author[0000-0001-9330-5003]{Pen\'{e}lope~Longa-Pe\~{n}a}
\affiliation{Centro de Astronom\'{i}a, Universidad de Antofagasta, Avenida Angamos 601, Antofagasta 1270300, Chile}

\author[0000-0001-9647-2886]{Jerome A. Orosz}
\affiliation{Astronomy Department, San Diego State University, 5500 Campanile Drive, San Diego, CA 92182, USA}

\author[0000-0003-1621-9392]{Mark T. Reynolds}
\affiliation{Department of Astronomy, University of Michigan, 1085 South University Avenue, Ann Arbor, MI 48109, USA}


\begin{abstract}
The optical light curves of quiescent black hole low-mass X-ray binaries often exhibit significant non-ellipsoidal variabilities, showing the photospheric radiation of the companion star is veiled by other source of optical emission. Assessing this ``veiling" effect is critical to the black hole mass measurement. 
Here in this work, we carry out a strictly simultaneous spectroscopic and photometric campaign on the prototype of black hole low-mass X-ray binary \azero.
We find that for each observation epoch, the extra optical flux beyond a pure ellipsoidal modulation is positively correlated with the fraction of  veiling emission, indicating the accretion disk contributes most of the non-ellipsoidal variations. Meanwhile, we also obtain a K2V spectral classification of the companion, as well as the measurements of the companion's rotational velocity $v \sin i = 83.8\pm1.9$~km~s$^{-1}$ and the mass ratio between the companion and the black hole $q=0.063\pm0.004$. 
\end{abstract}

\keywords{black hole physics --- stars: black holes --- X-rays: binaries}

\section{Introduction}

Mass and spin can fully describe an astrophysical black hole. Accurate measurements of these two parameters are fundamental to the studies of black hole properties, as well as a variety of astrophysical topics, e.g., supernova progenitors, binary evolution, and jet launching mechanisms (see reviews of \citealt{Casares2014,McClintock2014,Reynolds2021}). Stellar-mass black holes have long been identified in \xray\ binaries \citep{Remillard2006,Corral2016,Tetarenko2016}. Recently, large stellar spectroscopic surveys in the optical and infrared bands \citep[e.g.,][]{Liu2019,Thompson2019} and gravitational wave detections \citep{Abbott2021} provide novel approaches of discovering stellar-mass black holes. 

The mass of the black hole in a binary system can be measured using the dynamical method via the following equation: 
\begin{eqnarray}
f(M) \equiv \frac{PK_{\rm c}^3}{2\pi G} = \frac{M_{\rm BH}\sin^3 i}{(1+q)^2},
\label{eqn:mass}
\end{eqnarray}
where $P$ is the orbital period of the binary, $K_{\rm c}$ is the semi-amplitude of the radial velocity (RV hereafter) curve of the companion star, $i$ is the systemic inclination angle, and $q\equiv M_{\rm c}/M_{\rm BH}$ is the mass ratio between the companion and the black hole. The black hole \xray\ binaries with low-mass companions (hereafter BH LMXBs) have mass transfers via the Roche-lobe overflow mechanism. The optical/infrared (OIR) emission from the companion stars is detectable during the \xray\ quiescent state. In this case, the black hole mass can be dynamically measured via OIR spectroscopy and photometry. For the majority of BH LMXBs, the most challenging step is constraining the systemic inclination $i$. The thermal continuum from the photosphere of the Roche-lobe filling companion star should exhibit ellipsoidal modulation, based on which the systemic inclination can be derived. 
However, even during the \xray\ quiescence, the OIR light curves often show significant additional variabilities beyond ellipsoidal modulations \citep[e.g.,][]{Zurita2003,Shahbaz2004,Reynolds2007}, especially in the {\it active} optical state \citep{Cantrell2008}. \citet{Zurita2003} and \citet{Hynes2003} examined the short-timescale flaring variabilities in the optical band for BH LMXBs and their potential physical mechanisms. They concluded that these variabilities mostly originate from the accretion disk, i.e., the photospheric emission of the companion star is often {\it veiled} by the accretion-disk contribution. 

Contribution of the veiling emission could be high ($\simgt 40\%$) and variable, which cannot be ignored for the purpose of measuring black hole mass (see detailed discussions in \citealt{Wu2015,Wu2016}). A key approach to assess the veiling effect is to perform simultaneous OIR spectroscopy and photometry. The fraction of veiling emission can be measured from the spectroscopy and then be directly compared to the non-ellipsoidal variability. \citet{Wu2015,Wu2016} carried out the first such observational campaign on a BH LXMB system Nova Muscae 1991. Although the veiling fraction measurement for each individual spectrum suffers from large uncertainty ($\sim10\%$) mainly due to the faintness of the object, the average veiling fraction over the full orbital phase is instrumental in obtaining reliable constraints on the systemic inclination $i$ and black hole mass $M_{\rm BH}$. Here in this work, we utilize the simultaneous optical spectroscopy and photometry of the prototypical BH LMXB \azero\ to study the relationship between the veiling fraction and non-ellipsoidal variations exhibited in the light curve. 

\azero\ is the first identified BH LMXB \citep{McClintock1986}. A variety of dynamical studies have since been performed on this system (see Table~1 of \citealt{Casares2014} and the references therein). The orbital period is $\approx7.75$ hr. The spectral classifications of the companion in previous studies are not well consistent, spanning from K2V to K7V. A broad range of systemic inclination $i$ ($\sim30$--$75^\circ$) results in large uncertainty for the black hole mass ($\sim3$--14 $M_{\odot}$). \citet{Cantrell2010} selected the {\it passive}-state OIR light curves when veiling fraction was at minimum. They constrained the inclination $i=51.0^{\circ}\pm0.9^{\circ}$ and therefore the black hole mass $M_{\rm BH} = 6.61 \pm 0.25\ M_{\odot}$. On the other hand, \citet{Grunsven2017} obtained $i=54.1^{\circ}\pm1.1^{\circ}$ and $M_{\rm BH} = 5.86 \pm 0.24\ M_{\odot}$ based on the same photometric datasets as those in \citet{Cantrell2010}, but with different light curve models and fitting software. They argued that such small errorbars of inclination $i$ are underestimated, and the disk veiling effect may generate systematic uncertainties larger than the statistical ones. They further emphasized the importance of simultaneous OIR spectroscopy and photometry for reliable measurements of black hole mass. 

Our optical spectroscopic and photometric campaigns of \azero\  (see Section~\ref{data}) only cover $\sim 1/7$ of a full orbital period. The goal of this work is not a full dynamical modeling of this system. Rather, we provide a reliable spectral classification of the companion and measure its RV (Section~\ref{rv}). Then we measure the rotational broadening of the companion (thus the mass ratio $q$) and the fraction of veiling emission to the optical spectra $f_{\rm veil}$, for which we call ``veiling factor" (Section~\ref{vsini}). With that, we investigate the relation between the disk veiling effect and the non-ellipsoidal variations in the light curve (Section~\ref{relation}). \azero\ is $\sim2$ magnitude brighter in the $V$ band than Nova Muscae 1991. The reduced uncertainties of veiling measurements make it possible to perform such studies for each individual spectrum and its corresponding photometric magnitude. Section~\ref{summary} is a summary of our results. 

\section{Observation AND Data Reduction}\label{data}

The optical spectroscopy and photometry of \azero\ were obtained during the same observation run for Nova Muscae 1991 (see Section 2 of \citealt{Wu2015}). 
Table~\ref{tab:obs} lists the observation log. For the first observation epoch in the sequence, the spectroscopy and photometry started $\approx7.5$ min apart. They also have different exposure times (180 vs. 60 sec). For all the following epochs, the spectroscopy and photometry started at the same or nearly the same time (within 3 sec). They have exactly the same exposure lengths. We will include all the twelve epochs in the spectral analysis in Sections \ref{rv} \& \ref{vsini}, while the first one will be excluded from the analysis of the relation between veiling emission and photometric variability due to its non-simultaneity. 

The spectra of \azero\ were taken with the moderate-resolution echellette spectrograph MagE \citep{Marshall2008} mounted on the Magellan/Clay Telescope (moved to Magellan/Baade in 2015) at the Las Campanas Observatory. The spectrograph generates 15 orders (\#6--20) of spectra, covering 3100--11000~\AA\ in total. The wavelength dispersion ranges from 0.2~\AA\ (order \#20) to 0.7~\AA\ (order \#6) per pixel, while the velocity dispersion remains 22 km~s$^{-1}$.  The chosen slit width was $0.85^{\prime\prime}$, resulting in a spectral resolution of $\sim5000$. Alongside the spectra of \azero\ and Nova Muscae 1991, 70 spectra of 38 standard stars (with spectral types K0 to K7) were taken as the RV templates. Reduction of the MagE data includes bias correction, flat-fielding and wavelength calibration, which were performed with the MagE pipeline developed by Carnegie Observatories\footnote{https://code.obs.carnegiescience.edu/mage-pipeline} (for further details, refer to Section~2.1 of \citealt{Wu2015}). 

\begin{center}
\begin{deluxetable}{cccc}
\tabletypesize{\footnotesize}
\tablecaption{The Observation Log of \azero \label{tab:obs}}
\tablewidth{0pt}
\tablehead{ \colhead{Seq. No.} &  \colhead{Observation Start} & \colhead{Exposure Time} \\ 
\colhead{} & \colhead{(UT)} & \colhead{(sec)} 
}
\startdata
	                      & \multicolumn{1}{c}{2009 Apr 24} & \\
	    1                & 23:04:56/23:12:31\tablenotemark{a} &   180/60\tablenotemark{a} \\
	    2                & 23:17:50/23:17:47\tablenotemark{a} &  300 \\		
	    3                & 23:25:53 &  300  \\
	    4                & 23:32:46/23:32:45\tablenotemark{a} & 240 \\
	    5                & 23:38:30 &  240 \\
	    6                & 23:44:10/23:44:11\tablenotemark{a} &  240 \\
	    7                & 23:49:50 &  240 \\
	    8                & 23:55:20 & 240 \\
	    \hline
	    & \multicolumn{1}{c}{2009 Apr 25} & \\
	    9                & 00:00:55 &  240 \\	
	    10               & 00:06:30 &  240 \\
	    11               & 00:12:00 & 240 \\
	    12               &  00:17:30 &  240 \\	
\enddata
\tablenotetext{a}{The values for the spectroscopic (former) and photometric (latter) observations, respectively. For other entries, the values are exactly the same for both spectroscopic and photometric observations.}
\end{deluxetable}
\end{center}

Photometric observations of \azero\ in the Johnson $V$ band were carried out with the 2.5-meter du Pont Telescope also located at the Las Campanas Observatory. The $2048\times2048$ Tek\#5 CCD camera covers a field of view of $8.85^\prime \times 8.85^\prime$ with the pixel size of $0.259^{\prime\prime}$. Standard {\sc iraf} procedures were performed for bias correction, flat-fielding, and aperture photometry. The aperture size is 1.5 times the FWHM of the reference stars, which were chosen from nearby non-variable stars. We selected four reference stars that provide the most consistent photometric calibration. The final adopted $V$-band magnitudes of \azero\ are the average of the results from these four references. Dereddening is not necessary for this work because neither color information nor absolute magnitude is involved. 

\begin{center}
\begin{deluxetable*}{cccccccc}
\tabletypesize{\footnotesize}
\tablecaption{Spectral Analysis Results\label{tab:orders}}
\tablewidth{0pt}
\tablehead{ \colhead{Order \#} & \colhead{Central $\lambda$} & \colhead{utilized $\lambda$ range} &\colhead{$K_{\rm c}$} &
   \colhead{$u$ (limb} & \colhead{\vsini} &
    \colhead{$f_{\rm star}$} & \colhead{$f_{\rm veil}$} \\
\colhead{} & \colhead{(\AA)} & \colhead{(\AA)} &\colhead{(km~s$^{-1}$)} &
   \colhead{ darkening)} & \colhead{(km~s$^{-1}$)} &
  \colhead{(\%)} &  \colhead{(\%)}
  }
\startdata
12 & $5140$ & 4910--5400 & $426.8\pm27.6$ & $0.795$ &
$83.5\pm2.4$ & $62.1\pm7.0$ & $37.9\pm7.0$ \\
11 & $5590$ & 5350--5560, 5590--5775, 5790--5870 & $437.0\pm31.6$ & $0.747$ &
$84.0\pm2.5$ & $66.2\pm7.2$ & $33.8\pm7.2$ \\
\hline
average & & & & & $83.8\pm1.9$ & $64.2\pm6.6$ & $35.8\pm6.6$\\
\enddata
\tablecomments{The quoted uncertainties are at the $1\sigma$ level of
  confidence.}  
\end{deluxetable*}
\end{center}

\section{Radial Velocity Analysis}\label{rv}

The order \#12 MagE spectra are expected to provide the best quality results since the wavelength range ($\sim4700$--5500 \AA) contains strong and abundant photospheric absorption lines of K dwarfs, with minimal contaminations from telluric and interstellar features. This has been verified in the work of Nova Muscae 1991 \citep{Wu2015}. We start our investigations with the order \#12 spectra. Meanwhile, the order \#11 spectra ($\sim5170$--6000 \AA) are also analyzed since these two orders combined cover most of the Johnson $V$-band filter curve. For each order, we masked out the accretion-disk related emission lines, telluric and interstellar features, as well as $\sim100$~\AA\ regions at both ends. Table~\ref{tab:orders} lists the central wavelength, the utilized wavelength ranges, and the main spectral measurements for both orders. 

\begin{figure*}
\begin{center}
\includegraphics[angle=0,width=.90\linewidth]{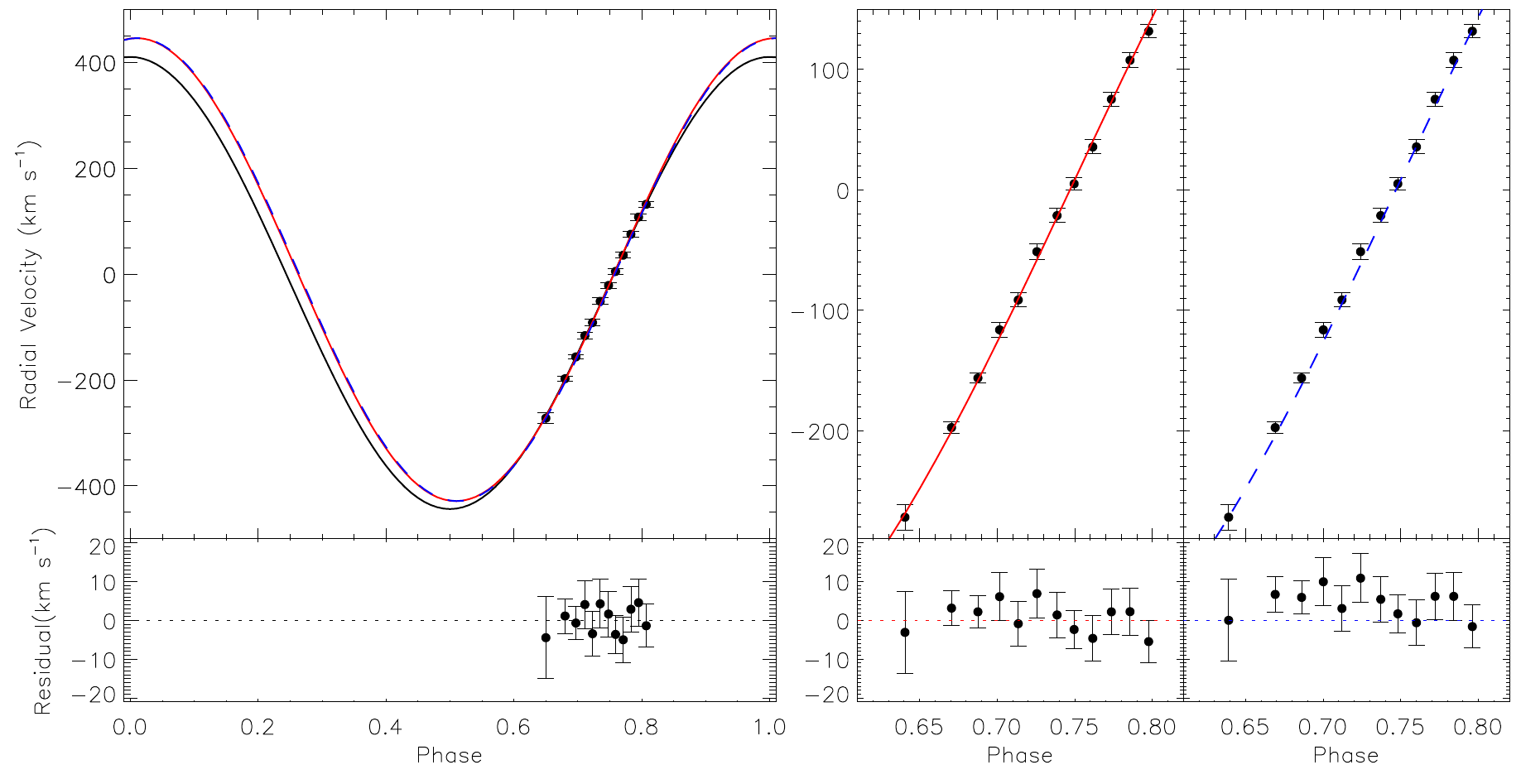}
\caption{\footnotesize Our RV measurements for the order \#12 spectra (filled black circles with error bars) and our best-fit RV curve (black solid lines). The red solid line represents the RV curve in \citet{Gonzalez2010}. We also re-calculated this curve after considering the orbital decay of \azero\ (blue dashed line). The right part is a zoom-in of the orbital phase range $\sim 0.6$--0.8. All the lower panels illustrate the residuals between the RV measurements and the corresponding curve.}\label{fig:rv}
\end{center}
\end{figure*}

\subsection{Spectral Classification of the Companion Star}\label{sec:cross}

The RV of the companion star is measured by cross-correlating \azero\ spectra to standard stellar spectral templates. This is performed using the \verb+fxcor+ task in the {\sc iraf} package, which employs the algorithm developed by \citet{Tonry1979}. The \verb+fxcor+ task  returns a parameter $R$ (hereafter TDR) representing the signal-to-noise ratio of the cross-correlation. We adopt the same method as in \citet{Wu2015} to select the best RV template from the 70 standard star spectra taken in the same observing run, i.e., the template spectrum returning the maximum TDR value when being cross-correlated with the average \azero\ order \#12 spectra (velocity-shifted to a uniform rest frame before averaging). The best-selected template spectrum is that of HD~136713 (TDR = 83.9), which is classified as a K2V star in the {\sc simbad} database \citep{Wenger2000}. This represents the spectral classification of the companion star in \azero. 

In order to confirm this spectral classification, we perform an independent check with the PHOENIX stellar spectral library \citep{Allard1995,Husser2013}. This library contains synthetic spectra generated by the PHOENIX stellar atmosphere code \citep{Hauschildt1999}. The covered stellar parameter ranges are effective temperature 2300~K $\leqslant T_{\rm eff} \leqslant $ 12000~K, surface gravity 0.0 $\leqslant \log g \leqslant +6.0$, and metallicity $-4.0 \leqslant$ [Fe/H] $\leqslant +1.0$. We carry out the cross-correlation analysis between the average \azero\ order \#12 spectra and the synthetic spectra in the PHOENIX library (re-binned to match the spectral resolution) using the \verb+spectool+ package.\footnote{https://gitee.com/zzxihep/spectool} The stellar spectrum providing the highest cross-correlation function value (c.c.f = 0.81) has the parameters of $T_{\rm eff} = 5000(\pm100)$~K, $\log g = 5.0(\pm0.5)$, and [Fe/H] $= 0.0(\pm0.5)$.\footnote{The values in the brackets are the parameter steps for the PHOENIX library spectra in these ranges.} For the best-selected standard star HD~136713, the up-to-date measurements retrieved from {\sc simbad} show that $T_{\rm eff}=4911$--5142~K, $\log g = 4.56$--4.69, and [Fe/H] $= 0.02 $--0.17 \citep{Delgado2017,Aguilera2018,Luck2018,Soto2018}, which are well consistent with those of the best-matching synthetic spectrum in PHOENIX. Therefore, we confirm the K2V spectral classification for the companion star in \azero. We will use the template spectrum of HD~136713 in the subsequent analyses, as it was obtained with exactly the same instrument configuration in the same observing run as those of the \azero\ spectra. Systematic uncertainties related to the template choice will be discussed in Section~\ref{vsini:sys}. 






\subsection{Radial Velocity Measurements}

We cross-correlate each of the twelve \azero\ spectra to the HD~136713 template with the \verb+fxcor+ task and obtain their heliocentric RV values. We then attempt to fit the RV curve represented by the following equation, 
\begin{eqnarray}
V(t) = \gamma + K_{\rm c}\cos (2 \pi \frac{t-T_0}{P}),  
\end{eqnarray}
where $\gamma$ is the heliocentric systemic RV, and $t$ is the mid-point of the spectroscopic epoch in heliocentric Julian Days (HJD). Here we define phase zero ($T_0$) at the time of maximum RV. For the orbital period $P$, we first adopt the value of $P=0.32301405(1)$ days from \citet{Gonzalez2010}. Then we consider the orbital decay of \azero\ ($\dot{P} = -0.53\pm0.07\ \mu$s per orbital cycle) found by \citet{Gonzalez2014} and derive the value of $P=0.32301398(7)$ days at the time of our observation campaign. We obtained the same fitting results in both cases because of the minimal difference between the two period values.  

The best-fit $K_{c}$ values for order \#12 and order \#11 spectra are listed in Table~\ref{tab:orders}, which are consistent with previous works (e.g., $435.4\pm0.5$~km~s$^{-1}$ in \citealt{Neilsen2008} and $437.1\pm2.0$~km~s$^{-1}$ in \citealt{Gonzalez2010}). The error bars of our results are significantly larger because of the limited number of spectra (twelve) and phase coverage ($0.650$ -- 0.807). The purpose of the RV analysis here is not to better constrain the mass function. Our goal is to obtain reliable velocities for each spectrum (relative velocity to the template spectrum and the heliocentric velocity). These spectra will then be Doppler-shifted to the same reference frame as needed in subsequent analyses. This is verified by comparing our measurements and the RV curves in previous works. In the left half of Figure~\ref{fig:rv}, we plot our best-fit RV curve (black solid line) for order \#12, as well as the RV measurements for the twelve spectra and their residuals (lower panel). Also overplotted are the RV curve (red solid line) in \citet{Gonzalez2010} and that after considering the orbital decay mentioned above (blue dashed line). These two curves mostly overlap with each other. We present the zoomed-in comparison between these two curves and our RV measurements, which are clearly well consistent. The residuals are all $\simlt 1.5\sigma$. 


\section{Rotational Broadening and Veiling Measurements}\label{vsini}

For accreting BH LMXBs like \azero, the photospheric absorption lines from the companion are broadened because the system is tidally locked and hence the rotational velocity \vsini\ of the companion is significantly higher than that if it were a field star (typically a few km~s$^{-1}$). We utilize the {\it optimal subtraction} technique \citep{Marsh1994} to measure the rotational velocity \vsini\ and the fraction of contribution from the companion star $f_{\rm star}$. Then we can obtain the mass ratio $q$ using the formula $v\sin i/K_{\rm c} = 0.462q^{1/3}(1+q)^{2/3}$ \citep{Wade1988} which is applicable to Roche-lobe overflow accreting systems, as well as the veiling factor $f_{\rm veil} = 1 - f_{\rm star}$. 

\begin{figure}
\includegraphics[angle=0,width=.99\linewidth]{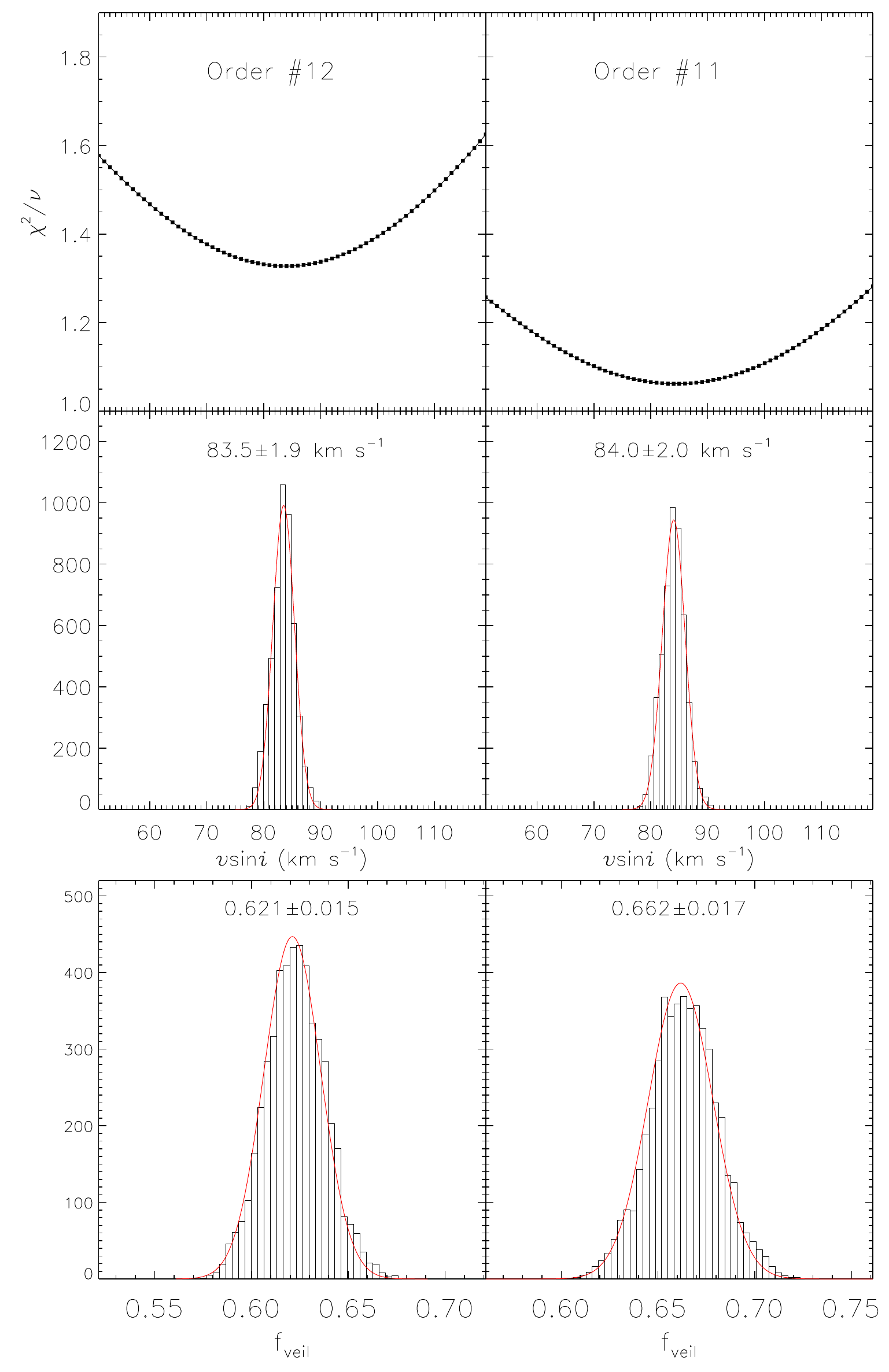}
\caption{\footnotesize Measurement of the companion rotational velocity \vsini\ and the veiling factor $f_{\rm veil}$ for orders \#12 (left) and \#11 (right). The upper panels present the reduced $\chi^2$ curve for the trial velocity values. The degree of freedom $\nu$ is 1268 and 1125 for orders \#12 and \#11, respectively. The middle and lower panels show the statistical distributions of \vsini\ and $f_{\rm veil}$ obtained with the bootstrap procedure. The red solid lines represent the best-fit Gaussian profiles to the histograms. The measured \vsini\ and $f_{\rm veil}$ values and their statistical uncertainties for orders \#12 and \#11 are labeled in each of the middle and lower panels. }\label{fig:vsini}
\end{figure}

The {\it optimal subtraction} procedure is carried out using the {\sc molly} software developed by T.~R.~Marsh.\footnote{http://deneb.astro.warwick.ac.uk/phsaap/software/molly} The template spectrum of HD~136713 is first broadened with a certain rotational velocity and multiplied by the companion contribution factor $f_{\rm star}$. Then it is subtracted from the average spectrum of \azero, which is already shifted to the rest frame of the template spectrum. The residual spectrum is examined with a $\chi^2$ test to see whether the stellar absorption features have all been subtracted. The minimum $\chi^2$ value determines the measured rotational velocity \vsini, and stellar contribution $f_{\rm star}$. We adopted a velocity trial grid from 50 to 120 km~s$^{-1}$ in steps of 1 km~s$^{-1}$. The resulting $\chi^2$ curves for orders \#12 and \#11 spectra are shown in the upper panels of Figure~\ref{fig:vsini}. 

During this process, the template spectrum of HD~136713 has been smeared based on the phase and exposure time of each \azero\ MagE spectrum to account for the velocity smearing during the spectroscopy. The limb darkening effect has also been properly considered. We calculated the 
limb darkening coefficient $u$ for both orders \#12 and \#11 by linearly interpolating their central wavelengths (see Table~\ref{tab:orders}) to those of the $B$ and $V$ bands, for which the linear limb darkening coefficients are retrieved from \citet{Claret2013}, assuming $T_{\rm eff} = 5000$~K, $\log g=4.5$, and [Fe/H] = 0 (which are the grid values closest to the stellar parameters of HD~136713).  The adopted limb darkening coefficient is 0.795 for order \#12, and 0.747 for order \#11. 

In order to estimate the uncertainties of \vsini\ and $f_{\rm star}$ measurements, we generate 5000 simulated spectra of \azero\ using the bootstrap method (see details in, e.g.,  \citealt{Steeghs2007,Wu2015}). We repeat the optimal subtraction procedure for each simulated spectrum. The statistical distributions of the measured \vsini\ and $f_{\rm star}$ values from 5000 simulated spectra can both be described by a Gaussian profile (see the middle and lower panels of Figure~\ref{fig:vsini}). The mean of the \vsini\ Gaussian distribution is well consistent with the trial velocity resulting in the minimum $\chi^2$ value (within 1 km~s$^{-1}$). For both  \vsini\ and $f_{\rm star}$, we adopt the mean of the best-fit Gaussian profile as the measured value and the standard deviation as the statistical uncertainty. The rotational velocities are $83.5\pm1.9$~km~s$^{-1}$ and $84.0\pm2.0$~km~s$^{-1}$, while the companion star contributions $f_{\rm star}$ are $(62.1\pm1.5)\%$ and $(66.2\pm1.7)\%$ for orders \#12 and \#11, respectively. The veiling factor is calculated as $f_{\rm veil}= 1 - f_{\rm star}$, which is $(37.9\pm1.5)\%$ and $(33.8\pm1.7)\%$ for orders \#12 and \#11, respectively.

\subsection{Assessing the Systematic Uncertainties}\label{vsini:sys}

The effective temperature $T_{\rm eff}$ and metallicity [Fe/H] of the template spectrum will impact the strength of the photospheric line features. Therefore, the systematic uncertainties in the \vsini\ and $f_{\rm star}$ measurements related to template choice need to be assessed. For this purpose, we repeat our analyses with the \# order 12 spectra of five other templates listed in Table~\ref{tab:sys}, covering spectral type K1V to K4V. These template spectra were also obtained in the same observing run with the same instrument settings as those of the \azero\ spectra. They all exhibit good cross-correlations with  \azero\ spectra (TDR $> 60$) during the template-matching procedure described in Section~\ref{sec:cross}. All of these five standard stars, along with the chosen template star HD~136713, have $T_{\rm eff}$ and [Fe/H] measurements in \citet{Aguilera2018}, as listed in Table~\ref{tab:sys}. The typical measurement errors are $\sim 100$~K for $T_{\rm eff}$, and $\sim0.1$ for  [Fe/H]. These five template stars have effective temperature difference of $\sim -300$ to $+150$~K, and metallicity difference of $\sim -0.2$ to $+0.1$, as relative to HD~136713. The linear limb darkening coefficient $u$ for these templates are also calculated with the databases in \citet{Claret2012,Claret2013}. 

For the rotational velocity \vsini, all the other five templates yield similar values (differences $<1.5$~km~s$^{-1}$; within $1\sigma$) as that of HD~136713, indicating that the choice of template does not significantly impact the \vsini\ measurement, as also shown in \citet{Steeghs2007}. We take the standard deviation of the six measurements, which is  $0.5$~km~s$^{-1}$, as the systematic uncertainty of  \vsini\ measurements. We add it on top of the statistical uncertainty obtained previously using the bootstrap method. Therefore, the final adopted rotational velocities for \#12 and \#11 spectra are $83.5\pm2.4$~km~s$^{-1}$ and $84.0\pm2.5$~km~s$^{-1}$, respectively. The average of the two is $83.8\pm1.9$~km~s$^{-1}$. This value is well consistent with previous studies (e.g., $82\pm2$~km~s$^{-1}$ in \citealt{Neilsen2008}; $83\pm5$~km~s$^{-1}$ in \citealt{Marsh1994}, which is also adopted by \citealt{Gonzalez2010}). Using $K_{\rm c} = 437.1\pm2.0$~km~s$^{-1}$ from \citet{Gonzalez2010}, we obtain the mass ratio $q = 0.063\pm0.004$.  In this case, the mass of the black hole is $M_{\rm BH} = (3.16\pm0.05) \sin^{-3} i$~$M_{\odot}$, while that of the companion star is $M_{\rm c} = (0.20\pm0.02) \sin^{-3} i$~$M_{\odot}$. 

\begin{center}
\begin{deluxetable}{llrccc}
\tabletypesize{\footnotesize}
\tablecaption{Rotational Velocity and Veiling Factor for Different Templates (Order \#12)\label{tab:sys}}
\tablewidth{0pt}
\tablehead{ \colhead{Template} & \colhead{$T_{\rm eff}$} & \colhead{ [Fe/H] } & \colhead{\vsini } & \colhead{$f_{\rm veil}$}  \\
\colhead{} & \colhead{(K)} & \colhead{} & \colhead{(km~s$^{-1}$)} & \colhead{(\%)}  
}
\startdata
HD~136713 & $4947\pm100$ & $0.04\pm0.10$ & $83.5\pm1.9$ & $38.1\pm0.9$ \\
\hline
HD~170657 & $5102\pm50$ & $-0.14\pm0.04$ & $84.8\pm2.2$ & $29.3\pm2.0$ \\
HD~130992 & $4851\pm100$ & $-0.16\pm0.10$ & $83.6\pm2.0$ & $38.9\pm1.4$ \\
HD~31560  & $4704\pm100$ & $-0.10\pm0.10$ & $84.6\pm1.9$ & $41.9\pm1.2$ \\
HD~170493 & $4704\pm100$ & $0.11\pm0.10$ & $84.0\pm1.9$ & $44.4\pm1.1$ \\
HD~131977 & $4675\pm116$ & $0.05\pm0.04$ & $84.3\pm1.9$ & $43.7\pm1.1$ \\
\enddata
\tablecomments{The quoted uncertainties are at the $1\sigma$ level of
  confidence.}
\end{deluxetable}
\end{center}

Cooler stars generally have stronger photospheric absorption lines. Therefore, choosing a template spectrum with lower effective temperature will result in a higher veiling factor. Similarly, a higher-metallicity template will also lead to a higher veiling measurement. These are evident in our analysis (see the last column of Table~\ref{tab:sys}). HD~131977 has similar metallicity as that of HD~136713, while the effective temperature is $\sim300$~K lower; the veiling measurement is $\sim5\%$ higher. HD~31560 and HD~170493 have similar effective temperatures but the metallicities differ by $\sim0.2$ dex; the veiling value obtained with the metal-richer HD~170493 is $\sim3\%$ higher than that with HD~31560. HD~130992 has a lower effective temperature and lower metallicity than those of HD 136713; the effects on veiling measurements offset each other, and hence their values agree within $1\%$. In contrast, HD~170657 is hotter and metal-poorer than HD 136713; these two combined make the measured veiling value with HD~170657 $\sim9\%$ lower. We adopt the standard deviation of the six veiling values listed in Table~\ref{tab:sys} as a conservative estimate of the systematic uncertainty related to template selection, which is $5.5\%$. Again this uncertainty is added on top of the statistical uncertainty. The final adopted $f_{\rm veil}$ values for the two orders are listed in Table~\ref{tab:orders}. The order \#11 spectra covering redder wavelengths than the order \#12 spectra have a smaller veiling factor $f_{\rm veil}$, which is consistent with the findings of Nova Muscae 1991 and other similar systems (see \citealt{Wu2015} and the references therein). The average veiling factor over these two orders is $(35.8\pm6.6)\%$. This can be taken as the average disk contribution at the Johnson $V$ band during our simultaneous spectroscopy and photometry (phase 0.680--0.807). 

\subsection{Veiling Measurements for Individual Spectra}\label{vsini:indi}

\begin{center}
\begin{deluxetable}{lcccc}
\tabletypesize{\footnotesize}
\tablecaption{The Photometry and Veiling Measurements \label{tab:indi}}
\tablewidth{0pt}
\tablehead{ \colhead{Seq.} &  \colhead{Magnitude} & \colhead{$f_{\rm veil}$ (\%)} & \colhead{$f_{\rm veil}$ (\%)} & \colhead{$f_{\rm veil}$ (\%)}\\ 
\colhead{No.} & \colhead{($V$ band)} & \colhead{(order \#12)} & \colhead{(order \#11)} & \colhead{(average)}
}
\startdata
	    1\tablenotemark{a}                & $18.003\pm0.010$ &   $45.8\pm6.7$ & $27.0\pm8.0$ & $36.4\pm5.2$\\
	    2                & $17.989\pm0.004$ &  $33.2\pm3.4$ & $30.7\pm4.7$ & $31.9\pm2.9$ \\		
	    3                & $17.899\pm0.004$ &  $33.2\pm3.1$ & $35.4\pm5.2$ & $34.3\pm3.0$\\
	    4                & $17.914\pm0.004$ & $39.0\pm4.7$ & $34.2\pm6.1$ & $36.6\pm3.8$\\
	    5                & $18.058\pm0.004$ &  $32.7\pm4.3$ & $37.2\pm5.6$ & $34.9\pm3.5$\\
	    6                & $18.151\pm0.004$ &  $35.4\pm5.4$ & $22.6\pm6.9$ & $29.0\pm4.4$\\
	    7                & $18.085\pm0.004$ &  $25.2\pm5.4$ & $24.2\pm6.8$ & $24.7\pm4.3$\\
	    8                & $17.975\pm0.004$ & $41.9\pm4.4$ & $27.7\pm5.6$ & $34.8\pm3.6$\\
	    9                & $17.939\pm0.004$ &  $35.0\pm5.5$ & $40.3\pm6.5$ & $37.7\pm4.2$\\	
	    10               & $17.960\pm0.004$ &  $43.3\pm4.2$ & $39.7\pm6.5$ & $41.5\pm3.8$\\
	    11               & $17.958\pm0.004$ & $27.2\pm8.3$ & $36.4\pm6.5$ & $31.8\pm5.3$\\
	    12               &  $18.047\pm0.004$ &  $31.2\pm5.0$ & $33.9\pm6.7$ & $32.6\pm4.2$ \\	
\enddata
\tablenotetext{a}{This observation is not included in the analysis in Section~\ref{relation} because of the non-simultaneity of the spectroscopy and photometry. }
\end{deluxetable}
\end{center}

We then measure the veiling factor for each individual spectrum in order to investigate the relation between the veiling emission and optical variability. Each \azero\ spectrum is Doppler-shifted to the rest frame of the HD~136713 template spectrum. The velocity smearing of the template has also been implemented for each observation epoch. The rotational velocity of the companion would not vary significantly during the time span of our observations. We adopt the above \vsini\ value obtained with the average spectra for all the individual ones, which should be more reliable and has better precision than that measured with individual spectrum. In order to properly estimate the uncertainty of the veiling, we again apply the bootstrap method. We generate 5000 sets of spectra, each of which contains twelve simulated individual spectra. For each set, all the spectra are broadened with the same rotational velocity, which is randomly chosen from a Gaussian distribution defined by the \vsini\ value listed in Table~\ref{tab:orders} (e.g., for order \#12, $\mu=83.5$~km~s$^{-1}$,  $\sigma=2.4$~km~s$^{-1}$). The veiling factor $f_{\rm veil}$ for each simulated spectrum is obtained with the optimal subtraction technique. 

With these procedures, we have obtained 5000 veiling measurements for each individual epoch. Their statistical distribution can be well characterized with a Gaussian profile. The mean and standard deviation are adopted as the value and uncertainty of the veiling factor for each individual spectrum. The results for both orders are listed in Table~\ref{tab:indi}. The veiling uncertainties of order \#12 spectra are smaller than those of order \#11 spectra, although the latter generally have a higher signal-to-noise ratio. This is likely because the order \#12 spectra have more stellar photospheric spectral features which result in better constraints on the stellar contribution. We also calculate the average veiling factor over these two orders for each spectroscopic observation. The twelfth spectrum of \azero\ at phase 0.807 shows an average veiling factor of $(32.6\pm4.2)\%$. This is similar to the result of \citet{Cantrell2010} which gives a $V$-band veiling factor of $(35\pm3)\%$ at phase 0.804 (phase 0.554 in their definition). The systematic uncertainties related to template choosing (see Section~\ref{vsini:sys}) will uniformly increase or decrease the veiling measurements for all individual spectra. They are not added on top of the statistical uncertainty since we will focus on the variation of veiling in the following analysis. 

\section{The Relation between Veiling and Non-Ellipsoidal Variability}\label{relation}

With the simultaneous spectroscopic and photometric observations in this work, we can investigate the relation between the veiling emission and the non-ellipsoidal variability of accreting BH LMXBs like \azero. In this section, we exclude the first photometric and spectroscopic epochs because of their non-simultaneity. 

\begin{figure}[t]
\includegraphics[angle=0,width=.99\linewidth]{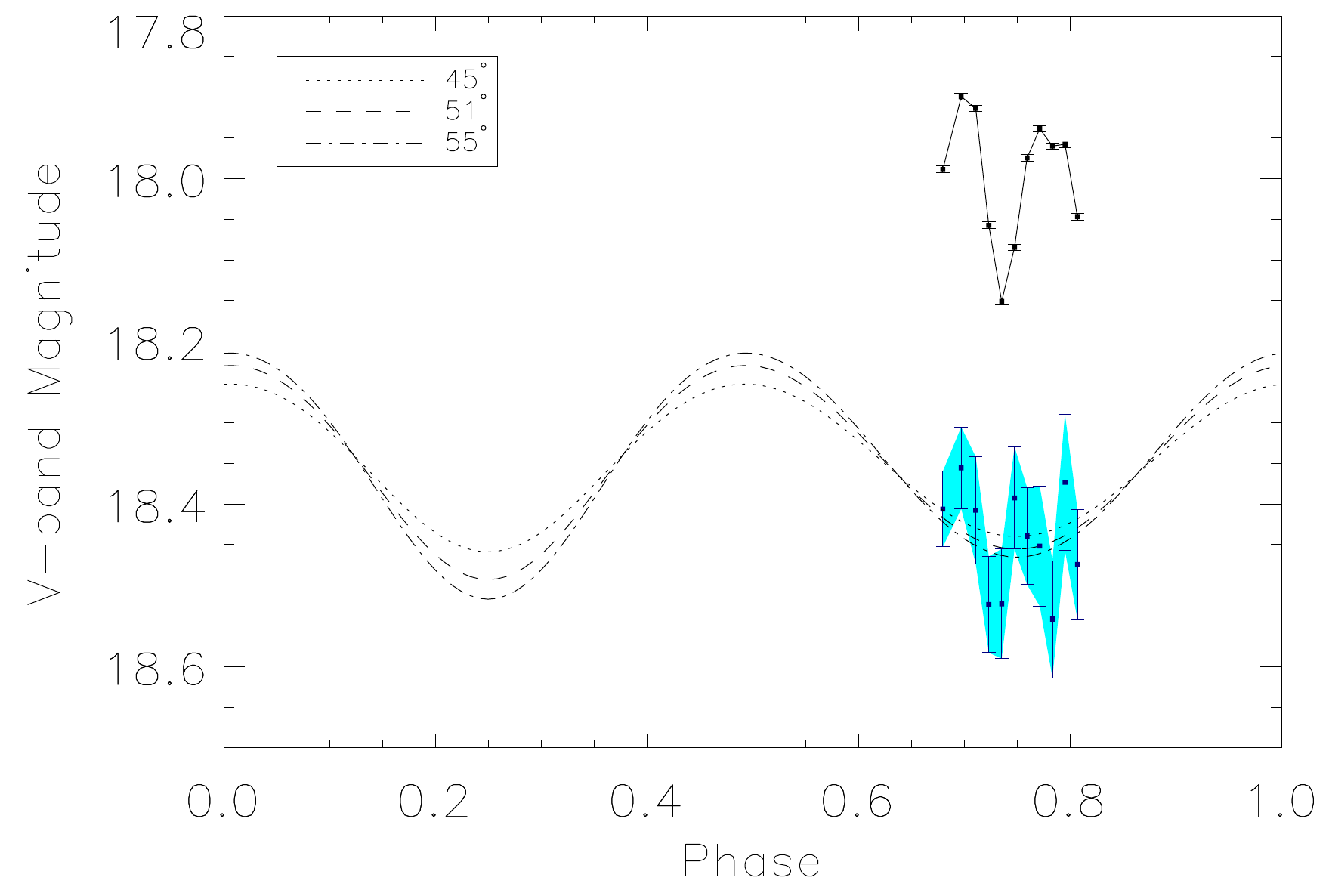}
\caption{\footnotesize Our $V$-band photometry of \azero\ (the upper-right black solid lines), as well as the pure ellipsoidal light curve generated by the ELC code with different inclination angles (see the legend). The blue symbols and cyan-shaded area show the veiling-corrected light curve and its uncertainty range.}\label{fig:lc}
\end{figure}

Figure~\ref{fig:lc} shows the observed $V$-band light curve (the upper right black solid line), which spans the magnitude range of $V_{\rm mag} = 17.90$--18.15. Although without $I$-band photometry, we are not able to directly use the criteria in \citet{Cantrell2008} to determine its optical state, \azero\ was likely in the {\it active} state during our observations based on its $V$-band brightness and variability amplitude. We correct the $V$-band photometry for the veiling effect, i.e., removing the veiling contribution and re-calculating the magnitudes. The corrected light curve is presented by the blue symbols and the cyan-shaded area showing its uncertainty range (see lower right part of Figure~\ref{fig:lc}). It could hint at an apparent quasi-periodicity at $\sim25$ min. However, this is only speculative given the limited phase coverage and number of data points. Also plotted are the ellipsoidal light curves purely arising from the Roche-lobe filling  companion star with the systemic inclination at $i = 45^\circ$ (dotted line), $51^\circ$ (the value from \citealt{Cantrell2010}; dashed line), and $55^\circ$ (dash-dotted line). These theoretical light curves are generated with the Eclipsing Light Curve code (ELC; \citealt{Orosz2000}); the dynamical parameters of the binary system and the stellar parameters of the companion are set to the values determined in this work. The $i = 51^\circ$ light curve is normalized so that the mean magnitude in the phase range 0.680--0.807 is the same as that of the veiling-corrected photometry data points which cover the same phases. The $i = 45^\circ$ and $i = 55^\circ$ light curves both have the same mean magnitude over a full orbital period as that of the $i = 51^\circ$ light curve. 

\begin{figure}[t]
\includegraphics[angle=0,width=.99\linewidth]{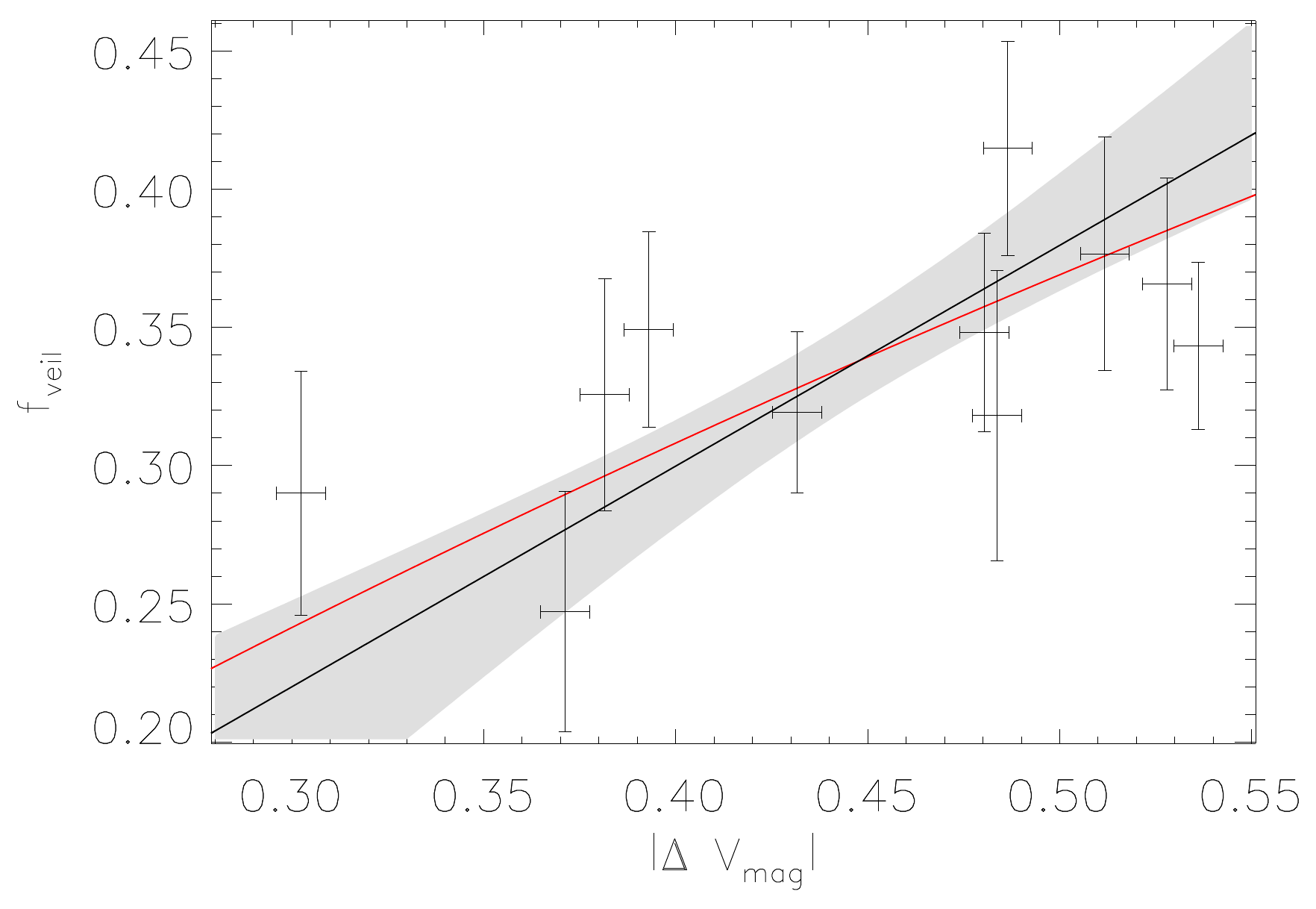}
\caption{\footnotesize The relation between the non-ellipsoidal brightness $|\Delta V_{\rm mag}|$ and the veiling factor $f_{\rm veil}$. The black solid line and the grey-shaded area represents the best-fit linear correlation and its 1$\sigma$ uncertainty range, while the red solid line shows the theoretical calculation. }\label{fig:veiling}
\end{figure}

The observed $V$-band light curve clearly shows non-ellipsoidal variabilities. We calculate the difference between the observed magnitude and that given by the theoretical $i = 51^\circ$ ellipsoidal light curve $|\Delta V_{\rm mag}|$ (i.e., the {\it extra} optical flux beyond the ellipsoidal modulation of the companion). We find that the magnitude difference is positively correlated with the veiling factor $f_{\rm veil}$ (see Figure~\ref{fig:veiling}), i.e., the epochs that the system is more brighter compared to pure ellipsoidal light curve do have a higher veiling factor. The Spearman's rank correlation coefficient is $\rho = 0.65$; the null-hypothesis probability is 0.029, corresponding to a statistical significance level of $2.2\sigma$. The best-fit linear relation between $|\Delta V_{\rm mag}|$ and $f_{\rm veil}$ is shown by the black solid line in Figure~\ref{fig:veiling}, while the grey-shaded area represent its $1\sigma$ uncertainty range. The linear relation is well consistent (within $1\sigma$) with the theoretical calculation (red solid line), i.e., $|\Delta V_{\rm mag}|= 2.5\log(1/f_{\rm star}) = 2.5\log(1/(1-f_{\rm veil}))$. 

The apparent scatter of the data points in Figure~\ref{fig:veiling} relative to the best-fit linear relation and the theoretical curve could be a combination of the measurement errors of $|\Delta V_{\rm mag}|$ and $f_{\rm veil}$, as well as the intrinsic scatter $\sigma_{\rm int}$. The latter represents the additional variabilities {\it not} related to the veiling. We apply a maximum likelihood approach to estimate $\sigma_{\rm int}$ relative to the theoretically calculated curve \citep{Kelly2007} and find that $\sigma_{\rm int}$ is consistent with zero. Therefore, the scatter can be fully explained by the measurement uncertainties of $|\Delta V_{\rm mag}|$ and $f_{\rm veil}$. As shown in Figure~\ref{fig:lc}, the veiling-corrected light curve is consistent with those generated from ellipsoidal modulation after considering the error bars.  However, the uncertainty ranges are substantially larger than the difference between light curves with $i=45^\circ$--$55^\circ$. We would argue that even using the individually veiling-corrected light curves it is still difficult to constrain the systemic inclination with desirable precision. Nevertheless, simultaneous spectroscopy and photometry on BH LMXBs covering full orbital cycles are still necessary to constrain the average veiling factor for given light curves, which is a critical input parameter for light curve modeling and reliable black hole mass measurement, as demonstrated by, e.g., \citet{Cantrell2010} and \citet{Wu2015,Wu2016}.

In essence, the {\it optimal subtraction} method we apply in this work measures the fraction of the photospheric emission of the companion star. Therefore, the ``veiling emission" we discuss here include all other sources of optical continuum. Nevertheless, it can shown that the accretion-disk emission should dominate the veiling continuum.  Chromospheric emission of the companion star may lead to some type of flaring activities. However, their luminosity alone is orders of magnitude lower than the observed optical luminosity \citep{Zurita2003}. Regarding to the non-stellar continuum emission, jets of quiescent black holes can contribute significant flickering variabilities in the OIR bands \citep[e.g.,][]{Dincer2018,Gallo2019}. However, this component follows a steep spectrum, i.e., stronger at redder optical and infrared bands \citep{Dincer2018}. Therefore, they will not dominate the $V$-band veiling emission studied in this work. \citet{Zurita2003} ruled out the possibility of stream-disk impact point as a major source of flickering variability because no correlation with orbital phase or systemic inclination is found. In contrast, the non-ellipsoidal variability are stronger for systems with lower-temperature companion stars, while the variability timescale positively correlates with the orbital period \citep{Zurita2003,Hynes2003}. These properties indicate that the veiling continuum is dominated by the accretion disk. \citet{Zurita2003} argued that the most likely physical mechanism is the magnetic loop reconnection occurred in the accretion disk. The reprocessed X-ray emission may also contribute. 

One caveat is that the spots in the photosphere of the companion could also produce the correlation between optical variability and veiling factor shown in Figure~\ref{fig:veiling}.  Indeed, starspots could generate up to 0.2 mag variability \citep{McClintock1990}. However, the timescales range from months to years \citep{Vogt1975,Bouvier1989}. As a comparison, \azero\ varied by $0.25$ mag in merely $\sim20$ minutes (see the 3rd and 6th epochs). We conclude that photospheric spots of the companion are not the cause of the positive correlation between non-ellipsoidal variability and the veiling factor.

\section{Summary}\label{summary}

In this work, we carry out simultaneous spectroscopic and photometric campaigns for the BH LMXB \azero. Although the limited number of observation epochs and phase coverage would not facilitate a full dynamical study, we obtain the K2V spectral classification of the companion, as well as its rotational broadening $v \sin i = 83.8\pm1.9$~km~s$^{-1}$ and the mass ratio for this system $q \equiv M_{\rm c}/M_{\rm BH}= 0.063\pm0.004$. 

We also measure the veiling factor $f_{\rm veil}$, i.e., the fraction of veiling continuum to the optical emission. The average veiling fraction at phase $0.650$--0.807 is $(35.8\pm6.6)\%$ for the wavelength range of $\sim 4900$--5900~\AA, which generally corresponds to the Johnson $V$ band. With the simultaneous spectroscopy and photometry, we find that for each individual epoch, the extra optical flux beyond pure ellipsoidal modulation is positively correlated with the veiling factor. The veiling continuum is dominated by the accretion disk. The non-ellipsoidal optical variabilities of BH LMXBs in \xray\ quiescent state are mostly generated by the accretion disk. 

\centerline{\\}

We thank the anonymous referee for the constructive comments and suggestions that improved the manuscript. We thank D. Steeghs, P. G. Jonker, and M. A. P. Torres for their contributions which are reported in \citet{Wu2015}. We thank T. Marsh for developing and sharing the {\sc molly} spectral analysis software. This work was supported by the National Science Foundation of China (U1938105, 11988101, 11925301), the National Key R\&D Program of China (2019YFA0405504, 2019YFA0405000), and the Strategic Priority Program of the Chinese Academy of Sciences (XDB41000000). J.Wu acknowledges support from the President Fund of Xiamen University (No. 20720190051). S.W. acknowledges support from the Youth Innovation Promotion Association of the CAS (id. 2019057). L.C.H. was supported by the National Science Foundation of China (11721303, 11991052) and the National Key R\&D Program of China (2016YFA0400702). P.L.P. was partly funded by “Programa de Iniciaci\'{o}n en Investigaci\'{o}n-Universidad de Antofagasta. INI-17-03". 

Support for the design and construction of the Magellan Echellette Spectrograph was received from the Observatories of the Carnegie Institution of Washington, the School of Science of the Massachusetts Institute of Technology, and the National Science Foundation in the form of a collaborative Major Research Instrument grant to Carnegie and MIT (AST0215989). This research has made use of the {\sc simbad} database, operated at CDS, Strasbourg, France.


\facility{Magellan:Clay (Magellan Echellette Spectrograph), du~Pont}


\begin{thebibliography}{99}

\bibitem[Abbott et al.(2021)]{Abbott2021} Abbott, R., Abbott, T.~D., Abraham, S., et al.\ 2021, \apjl, 913, L7. doi:10.3847/2041-8213/abe949

\bibitem[Aguilera-G{\'o}mez et al.(2018)]{Aguilera2018} Aguilera-G{\'o}mez, C., Ram{\'\i}rez, I., \& Chanam{\'e}, J.\ 2018, \aap, 614, A55. doi:10.1051/0004-6361/201732209

\bibitem[Allard \& Hauschildt(1995)]{Allard1995} Allard, F. \& Hauschildt, P.~H.\ 1995, \apj, 445, 433. doi:10.1086/175708

\bibitem[Bouvier \& Bertout(1989)]{Bouvier1989} Bouvier, J. \& Bertout, C.\ 1989, \aap, 211, 99

\bibitem[Cantrell et al.(2008)]{Cantrell2008} Cantrell, A.~G., Bailyn, C.~D., McClintock, J.~E., et al.\ 2008, \apjl, 673, L159. doi:10.1086/528793

\bibitem[Cantrell et al.(2010)]{Cantrell2010} Cantrell, A.~G., Bailyn, C.~D., Orosz, J.~A., et al.\ 2010, \apj, 710, 1127. doi:10.1088/0004-637X/710/2/1127

\bibitem[Casares \& Jonker(2014)]{Casares2014} Casares, J. \& Jonker, P.~G.\ 2014, \ssr, 183, 223. doi:10.1007/s11214-013-0030-6

\bibitem[Claret et al.(2012)]{Claret2012} Claret, A., Hauschildt, P.~H., \& Witte, S.\ 2012, \aap, 546, A14. doi:10.1051/0004-6361/201219849

\bibitem[Claret et al.(2013)]{Claret2013} Claret, A., Hauschildt, P.~H., \& Witte, S.\ 2013, \aap, 552, A16. doi:10.1051/0004-6361/201220942

\bibitem[Corral-Santana et al.(2016)]{Corral2016} Corral-Santana, J.~M., Casares, J., Mu{\~n}oz-Darias, T., et al.\ 2016, \aap, 587, A61. doi:10.1051/0004-6361/201527130

\bibitem[Delgado Mena et al.(2017)]{Delgado2017} Delgado Mena, E., Tsantaki, M., Adibekyan, V.~Z., et al.\ 2017, \aap, 606, A94. doi:10.1051/0004-6361/201730535

\bibitem[Din{\c{c}}er et al.(2018)]{Dincer2018} Din{\c{c}}er, T., Bailyn, C.~D., Miller-Jones, J.~C.~A., et al.\ 2018, \apj, 852, 4. doi:10.3847/1538-4357/aa9a46

\bibitem[Gallo et al.(2019)]{Gallo2019} Gallo, E., Teague, R., Plotkin, R.~M., et al.\ 2019, \mnras, 488, 191. doi:10.1093/mnras/stz1634

\bibitem[Gelino et al.(2001)]{Gelino2001} Gelino, D.~M., Harrison, T.~E., \& Orosz, J.~A.\ 2001, \aj, 122, 2668. doi:10.1086/323714

\bibitem[Gonz{\'a}lez Hern{\'a}ndez \& Casares(2010)]{Gonzalez2010} Gonz{\'a}lez Hern{\'a}ndez, J.~I. \& Casares, J.\ 2010, \aap, 516, A58. doi:10.1051/0004-6361/201014088

\bibitem[Gonz{\'a}lez Hern{\'a}ndez et al.(2014)]{Gonzalez2014} Gonz{\'a}lez Hern{\'a}ndez, J.~I., Rebolo, R., \& Casares, J.\ 2014, \mnras, 438, L21. doi:10.1093/mnrasl/slt150

\bibitem[Hauschildt \& Baron(1999)]{Hauschildt1999} Hauschildt, P.~H. \& Baron, E.\ 1999, Journal of Computational and Applied Mathematics, 109, 41

\bibitem[Husser et al.(2013)]{Husser2013} Husser, T.-O., Wende-von Berg, S., Dreizler, S., et al.\ 2013, \aap, 553, A6. doi:10.1051/0004-6361/201219058

\bibitem[Hynes et al.(2003)]{Hynes2003} Hynes, R.~I., Charles, P.~A., Casares, J., et al.\ 2003, \mnras, 340, 447. doi:10.1046/j.1365-8711.2003.06297.x

\bibitem[Kelly(2007)]{Kelly2007} Kelly, B.~C.\ 2007, \apj, 665, 1489. doi:10.1086/519947

\bibitem[Liu et al.(2019)]{Liu2019} Liu, J., Zhang, H., Howard, A.~W., et al.\ 2019, \nat, 575, 618. doi:10.1038/s41586-019-1766-2

\bibitem[Luck(2018)]{Luck2018} Luck, R.~E.\ 2018, \aj, 155, 111. doi:10.3847/1538-3881/aaa9b5

\bibitem[Marsh et al.(1994)]{Marsh1994} Marsh, T.~R., Robinson, E.~L., \& Wood, J.~H.\ 1994, \mnras, 266, 137

\bibitem[Marshall et al.(2008)]{Marshall2008} Marshall, J.~L., Burles, S., Thompson, I.~B., et al.\ 2008, \procspie, 7014, 701454. doi:10.1117/12.789972

\bibitem[McClintock et al.(2014)]{McClintock2014} McClintock, J.~E., Narayan, R., \& Steiner, J.~F.\ 2014, \ssr, 183, 295. doi:10.1007/s11214-013-0003-9

\bibitem[McClintock \& Remillard(1986)]{McClintock1986} McClintock, J.~E. \& Remillard, R.~A.\ 1986, \apj, 308, 110. doi:10.1086/164482

\bibitem[McClintock \& Remillard(1990)]{McClintock1990} McClintock, J.~E. \& Remillard, R.~A.\ 1990, \apj, 350, 386. doi:10.1086/168392

\bibitem[Neilsen et al.(2008)]{Neilsen2008} Neilsen, J., Steeghs, D., \& Vrtilek, S.~D.\ 2008, \mnras, 384, 849. doi:10.1111/j.1365-2966.2007.12599.x

\bibitem[Orosz \& Hauschildt(2000)]{Orosz2000} Orosz, J.~A. \& Hauschildt, P.~H.\ 2000, \aap, 364, 265

\bibitem[Remillard \& McClintock(2006)]{Remillard2006} Remillard, R.~A. \& McClintock, J.~E.\ 2006, \araa, 44, 49. doi:10.1146/annurev.astro.44.051905.092532

\bibitem[Reynolds(2021)]{Reynolds2021} Reynolds, C.~S.\ 2021, \araa, 59, 117. doi:10.1146/annurev-astro-112420-035022

\bibitem[Reynolds et al.(2007)]{Reynolds2007} Reynolds, M.~T., Callanan, P.~J., \& Filippenko, A.~V.\ 2007, \mnras, 374, 657. doi:10.1111/j.1365-2966.2006.11180.x

\bibitem[Shahbaz et al.(2004)]{Shahbaz2004} Shahbaz, T., Hynes, R.~I., Charles, P.~A., et al.\ 2004, \mnras, 354, 31. doi:10.1111/j.1365-2966.2004.08162.x

\bibitem[Soto \& Jenkins(2018)]{Soto2018} Soto, M.~G. \& Jenkins, J.~S.\ 2018, \aap, 615, A76. doi:10.1051/0004-6361/201731533

\bibitem[Steeghs \& Jonker(2007)]{Steeghs2007} Steeghs, D. \& Jonker, P.~G.\ 2007, \apjl, 669, L85. doi:10.1086/523848

\bibitem[Tetarenko et al.(2016)]{Tetarenko2016} Tetarenko, B.~E., Sivakoff, G.~R., Heinke, C.~O., et al.\ 2016, \apjs, 222, 15. doi:10.3847/0067-0049/222/2/15

\bibitem[Thompson et al.(2019)]{Thompson2019} Thompson, T.~A., Kochanek, C.~S., Stanek, K.~Z., et al.\ 2019, Science, 366, 637. doi:10.1126/science.aau4005

\bibitem[Tonry \& Davis(1979)]{Tonry1979} Tonry, J. \& Davis, M.\ 1979, \aj, 84, 1511. doi:10.1086/112569

\bibitem[van Grunsven et al.(2017)]{Grunsven2017} van Grunsven, T.~F.~J., Jonker, P.~G., Verbunt, F.~W.~M., et al.\ 2017, \mnras, 472, 1907. doi:10.1093/mnras/stx2071

\bibitem[Vogt(1975)]{Vogt1975} Vogt, S.~S.\ 1975, \apj, 199, 418. doi:10.1086/153705

\bibitem[Wade \& Horne(1988)]{Wade1988} Wade, R.~A. \& Horne, K.\ 1988, \apj, 324, 411. doi:10.1086/165905

\bibitem[Wenger et al.(2000)]{Wenger2000} Wenger, M., Ochsenbein, F., Egret, D., et al.\ 2000, \aaps, 143, 9. doi:10.1051/aas:2000332

\bibitem[Wu et al.(2015)]{Wu2015} Wu, J., Orosz, J.~A., McClintock, J.~E., et al.\ 2015, \apj, 806, 92. doi:10.1088/0004-637X/806/1/92

\bibitem[Wu et al.(2016)]{Wu2016} Wu, J., Orosz, J.~A., McClintock, J.~E., et al.\ 2016, \apj, 825, 46. doi:10.3847/0004-637X/825/1/46

\bibitem[Zurita et al.(2003)]{Zurita2003} Zurita, C., Casares, J., \& Shahbaz, T.\ 2003, \apj, 582, 369. doi:10.1086/344534


\end{thebibliography}
\end{document}